\documentclass[preprint,showpacs,preprintnumbers,amsmath,amssymb]{revtex4-1}
\usepackage{amsmath}
\usepackage{graphicx}
\usepackage{amssymb}
\usepackage{bm}

\begin{document}

\title{Multiple reflection-asymmetric type band structures in $^{220}$Th and dinuclear model}
\author{T.M. Shneidman}
\author{G.G. Adamian}
\author{N.V. Antonenko}
\author{R.V.Jolos}
\affiliation{Joint Institute for Nuclear Research, 141980 Dubna, Russia}
\author{W. Scheid}
\affiliation{Institut f\"ur Theoretische Physik der
Justus-Liebig-Universit\"at,
D--35392 Giessen, Germany}

\begin{abstract}
The negative parity bands in $^{220}$Th are analyzed within the dinuclear system model which was previously used for describing the alternating-parity bands in deformed actinides. The model is based on the assumption that cluster type shapes are produced by the motion of nuclear system in the mass-asymmetry coordinate. To describe the reflection-asymmetric collective modes characterized by  nonzero values of $K$,  the intrinsic excitations of clusters are taken into account.  The angular momentum dependence of the parity splitting and the staggering behavior of the  $B(E1)/B(E2)$ ratios as  functions of angular momentum are explained.
\end{abstract}

\date{Today}
\maketitle

\section{Introduction}
In the even-even isotopes of actinides and also in the heavy Ba and Ce isotopes the low-lying negative parity states are observed together with the usually presented collective positive-parity states combined into rotational or quasirotational ground-state bands. The formation of the positive-parity rotational or quasirotational  bands are connected in general to the
quadrupole collective motion, while the lowering of the negative-parity states is a signature of the presence of the reflection asymmetric collective mode. There are several approaches to treat the collective motion related to the reflection asymmetric degrees of freedom. One of them is based on the concept of the nuclear mean field \cite{nazar}
which has a static mirror asymmetric deformation or is characterized by a large amplitude of reflection asymmetric vibrations around the equilibrium shape. Another approach is based on the assumption that the reflection asymmetric shape is a consequence of the $\alpha$-clustering in nuclei \cite{iachello}. It is also known from the Nilsson-Strutinsky type calculations for light nuclei that nuclear configurations corresponding to the minima of the potential energy contain particular symmetries which are related to certain cluster structures \cite{rae,freer}.
Several calculations performed for heavy nuclei \cite{cwiok, aberg, shneyd} have shown that configurations with large equilibrium quadrupole deformations and low-lying collective negative parity states are strongly related to clustering.
We mention also a different approach to description of the properties of the alternating parity bands which is based on the idea of the aligned octupole phonons \cite{briancon,frauendorf}

The main idea of the cluster model developed in \cite {shneyd,shneyd1,shneyd2} is that a dynamics of a reflection asymmetric collective motion can be treated as a collective motion of nucleons between two clusters or as a motion in a mass-asymmetry coordinate. Such collective motion simultaneously creates deformations with even and odd-multipolarities. Among different cluster configurations only $\alpha$-cluster system  $^AZ \rightarrow ^{(A-4)}(Z-2)+^4$He gives a significant contribution to the formation of the low-lying nuclear states. Within this approach the existing experimental data on the angular momentum dependence of the parity splitting and multipole transition moments (E1,E2,E3) of the low-lying alternating parity states in odd and even actinides $^{220-228}$Ra, $^{223,225,227}$Ac,
$^{222-224,226,228-232}$Th, $^{231}$Pa, $^{232-234,236,238}$U and $^{240, 242}$Po and the medium mass nuclei
$^{144,146,148}$Ba, $^{151,153}$Pm, $^{146,148}$Ce $^{153,155}$Eu and $^{146,148}$Nd are well described. The perfect agreement between the calculated results and experimental data supplies the proof of the cluster features of reflection-asymmetric states. However, in our previous publications we have considered in the even-even nuclei only the low-lying collective negative parity states with $K$=0. However, there are  experimental data which indicate  a presence of the low-lying collective states related to the reflection-asymmetric modes which are characterized by  nonzero values of $K$. It can be also that  $K$ is not a good quantum number if nuclei are located in a transitional region between deformed and spherical ones. A good example is $^{220}$Th \cite{reviol} whose energy spectra is a challenge for the theoretical approaches.

To describe  the properties of the low-lying collective states related to the reflection-asymmetric collective mode and characterized by  nonzero values of $K$ in the framework of the cluster approach we should take into account intrinsic excitations of the clusters forming a nucleus under consideration. The aim of the present paper is a formulation of a simple model which gives a quantitative
description of  the low-lying positive and negative parity  bands in the spectra of nuclei like $^{220}$Th.
 Our model is based on the assumption that the collective oscillations of the
nuclear shape lead to the formation of the cluster-type configurations. This model is
further development of the previously used approaches \cite{shneyd,shneyd1,shneyd2}.

The choice of the collective coordinates and the procedure of the
calculation of the potential energy and of the inertia coefficients for the Hamiltonian of the model
are based on the concept of the dinuclear system (DNS). The concept of the dinuclear system was first introduced to explain the experimental data on deep inelastic and fusion reactions. Later on it was applied to the description of the nuclear structure phenomena, like alternating parity bands, mentioned above, and  superdeformed states \cite{Pb,Zn}.
The dinuclear system ($A$,$Z$) consists of two fragments ($A_1$,$Z_1$)
and ($A_2$,$Z_2$) with $A=A_1+A_2$ and $Z=Z_1+Z_2$ sticked closely together by a molecular-type
nucleus-nucleus potential. The  degrees of freedom needed to describe the
collective excitations of such a system are related to the rotation
of the DNS as a whole, to the relative motion of the DNS fragments,
to the intrinsic excitations of the fragments, and to the transfer
of nucleons between the DNS fragments. The latter process is
described usually by means of the mass-asymmetry coordinate
$\eta=(A_1-A_2)/(A_1+A_2)$ or by charge-asymmetry coordinate
$\eta_Z=(Z_1-Z_2)/(Z_1+Z_2)$.
The idea that the motion in mass-asymmetry is responsible for
the formation of reflection-asymmetric deformations of the medium
mass and heavy nuclei is based  on the observation which
manifest  the alpha-cluster dinuclear system for these nuclei.
Our calculations have shown that the potential energy for the alpha-cluster DNS are
close or even lower than the binding energy for these nuclei \cite{shneyd2}. This
is also in agreement with the fact that these nuclei are good
alpha-emitters.

Previously, we considered the mass-asymmetry motion variable   as the only intrinsic collective coordinate describing DNS. However,  the nuclei demonstrate yet another sign of
reflection-asymmetry, namely, low-lying (started below 1 MeV)
negative parity rotational bands with $K \ne$0 which can be related to the
non axially-symmetric reflection-asymmetric mode. The natural way to obtain such an excitation in the framework of the dinuclear system model is to consider
the relative angular motion of the  fragments  in the dinuclear systems illustrated in Fig. 1.
 It is the aim of the present investigation to extend the DNS model to
take into account such excitations.

Basically, there are two  physically different cases which  can be  distinguished. If the heavy fragment of DNS, whose lighter fragment is an $\alpha$-particle  is strongly deformed,  the light fragment  is tend to stay near the pole of the heavy cluster performing the small angular oscillations around this position. This case which holds for the heavy actinides is considered in the work []. If  the heavy fragment is nearly spherical, the interaction between the fragments is small to prevent the light cluster from rolling along  the surface of the heavy one. Such a case is realized in  the light isotopes of actinides, including the $^{220}$Th considered here. The structure of $^{220}$Th has been analyzed in several theoretical publications \cite{reviol,nazar1,otsuka}. Here, we apply an approach of the dinuclear system model.

\section{Model}
\subsection{Hamiltonian}

The degrees of freedom chosen to describe  a system with nearly spherical heavy cluster
are related to the rotation of the DNS as a whole,  to the quadrupole oscillations  of the heavy fragment, and to the transfer of nucleons between the fragments.
The Hamiltonian of the model can be presented in the form
\begin{eqnarray}
\hat H=\hat H_0+\hat V_{int},
\label{hamiltonian}
\end{eqnarray}
where $\hat H_0$ describes independent fragments of the  system and $\hat V_{int}$  describes the
interaction between the fragments.

%\begin{figure}
% Fig.1
%\vspace*{-2.0cm}
%\begin{center}
%\resizebox{0.80\textwidth}{!}{\includegraphics{{Fig3.eps}}}% Here is how to import EPS art
%\vspace*{-1.35 cm} \caption{\label{fig_1} Scheme of the experimental
%setup for the measurements of fission fragment angular
%distribution of binary reaction products. See explanations in the
%text for details.}
%\end{center}
%\end{figure}

\begin{figure}[t]
%\begin{right}
\includegraphics[bb=17 293 487 583, height=8.0 cm]{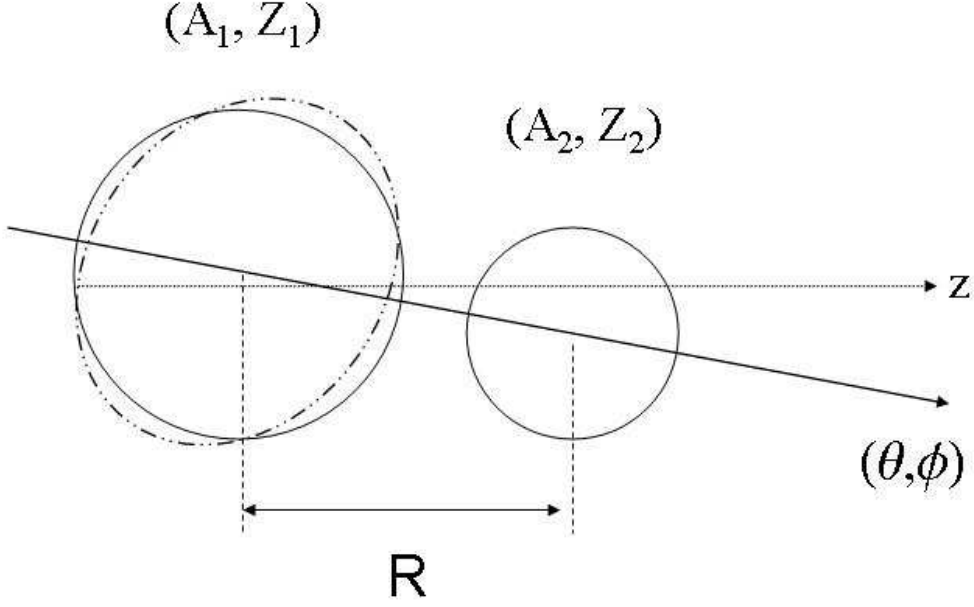}
%\end{right}
\begin{minipage}{16cm}
\vspace{0cm}
\caption{{\small \it
Schematic picture illustrates degrees of freedom used in the model to describe the dinuclear system. The orientation of the  vector of the relative distance $R$ are defined by the angles $\Omega (\theta,\phi)$ with respect to the laboratory frame system.}}
\end{minipage}
\label{fig1}
\end{figure}

We assume that the heavy cluster is spherical  and perform  harmonic quadrupole
oscillations around the spherically symmetric shape with frequency $\hbar \omega_0$,
 while the light cluster stays in its ground state. Then for
$\hat H_0$ we have the following expression
\begin{eqnarray}
\hat H_0= \hbar \omega_0 \hat n +\frac{\hbar^2}{2 \mu R^2}\hat L^2,
\label{h0}
\end{eqnarray}
where $\mu$ is the reduced mass of the DNS, $R$ is the distance between the centers of mass of the fragments, $\hat n$ is the operator of the number of the quadrupole phonons of heavy cluster and $\hat L^2$
is the operator of the square of angular momentum of relative rotations of the two fragments
\begin{eqnarray}
L^2= - \left [
\frac{1}{\sin \theta}\frac{\partial }{\partial \theta}
\sin \theta \frac{\partial }{\partial \theta}
+\frac{1}{\sin^2\theta} \frac{\partial^2}{\partial \phi^2}
\right ].
\end{eqnarray}

Angles $\Omega=(\theta,\phi)$ (see Fig.1) describes the orientation of the relative distance vector {\bf R} with
respect to the laboratory system.

Taking into account that for the low angular momenta the deformation of heavy fragment is small, the
interaction between the vibrational and rotational degrees of freedom can be taken in the lowest
order with respect to the operators of quadrupole deformation $\hat \beta_{2\mu}$
\begin{eqnarray}
\hat V_{int}=V_0 \sum_\mu \beta^*_{2\mu} Y_{2\mu}=
V_0 \beta_0 \left ( \left [d^++\tilde d \right ]\cdot Y_2\right ),
\end{eqnarray}
where we express the operator of quadrupole deformation through the creation and anihillation
operators of bosons
\begin{eqnarray}
\hat \beta_{2\mu}=\beta_0 (d^+_{2\mu}+\tilde d_{2\mu}),
\end{eqnarray}
with $\beta_0 =\sqrt{\hbar/2 B \omega_0}$.

In this work we does not consider the transfer of nucleons in the DNS explicitely. Instead, we assume that the system stays in its lowest state in mass-asymmetry coordinate. The reason for such an assumption is that the energy of $\alpha$-particle DNS is around 1 MeV higher than the binding energy of $^{220}$Th. The first excited state in mass-asymmetry will appear high enough to neglect its influence on the low-energy part of the spectra.
Instead of considering the mass-asymmetry motion we take the effective value of mass-asymmetry $\eta_0$ in our calculations. This value, which must be at least in the interval $\eta_\alpha<\eta_0<1$, where $\eta_\alpha=(A-4)/A$ is the mass-asymmetry for DNS with alpha-particle, is fixed by the description of the lowest 1$^-$ state.

If we neglect the interaction term $\hat V_{int}$ in (\ref{hamiltonian}) the eigenfunctions of the Hamiltonian
 can be constructed as
\begin{eqnarray}
\Psi^{IM}_{(n\tau n_\Delta I_1)I_2}= \left [ |n\tau n_\Delta I_1)\times Y_{I_2} \right ]_{(IM)},
\label{zerowf}
\end{eqnarray}
where $|n\tau n_\Delta I_1)$ represents the n-boson wave function of the heavy nucleus, with the seniority $\tau$
and angular momentum $I_1$. Since the quadrupole oscillations have positive parity, the   parity of the states
(\ref{zerowf}) is determined by the angular momentum of the relative rotation of the fragments $p=(-1)^{I_2}$. The energies of states
(\ref{zerowf}) are given in this approximation by the expresssion
\begin{eqnarray}
E_{n I_1 I_2 I}= \left [ \hbar \omega_0 n +\frac{\hbar^2}{2 \mu R^2}I_2(I_2+1) \right ].
\end{eqnarray}

The set of wave functions (\ref{zerowf}) can be used as a basis to construct the eigenfunction of the
Hamiltonian $\hat H$ in the form
\begin{eqnarray}
\Psi_{IM,p}=\sum_{I_1 I_2}\sum_{n_{I_1}\tau_{I_1}} a^{(I,p)}_{n_{I_1}\tau_{I_1}I_1I_2} \left [ |n_{I_1}\tau_{I_1}
I_1)\times Y_{I_2}\right ]_{(IM)}, \label{wf}
\end{eqnarray}
where coefficients $a^{(I,p)}_{n_{I_1}\tau_{I_1}I_1I_2}$ should be obtained by a diagonalization of $\hat H$. The
matrix elements of $\hat V_{int}$ between states (\ref{zerowf}) have the following form
\begin{eqnarray}
\langle\Psi^{IM}_{(n'\tau'I_1')I'_2}|\hat V_{int}|\Psi^{IM}_{(n\tau I_1)I_2}\rangle
&&= (-1)^{I_1+I'_2+I} V_0 \beta_0
\sqrt{\frac{5}{4\pi}(2I_2+1)} (I_2 0 2 0|I'_20)\nonumber \\
&& \times
 \left (
\begin{array}{c c c}
I'_2 & I'_1 & I \\
I_1 & I_2 & 2
\end{array}
\right ) (n'\tau'I'_1||(d^++\tilde d)||n \tau I_1),
 \label{matel}
\end{eqnarray}
where the reduced matrix elements of the boson operators can be calculated using  the  boson fractional parentage
coefficients
\begin{eqnarray}
&&\left [d^{n-1}(\alpha_1 I_1)dI|\}d^n\alpha I\right ]
=\frac{1}{\sqrt  n}\frac{1}{\sqrt{2I+1}}(d^n \alpha I ||d^+||d^{n-1}\alpha_1 I_1) \nonumber\\
&&\left [d^{n-1}(\alpha_1 I_1)dI|\}d^n\alpha I\right ] =(-1)^{I-I_1}\frac{1}{\sqrt  n}\frac{1}{\sqrt{2I+1}}(d^{n-1}
\alpha_1 I_1 ||\tilde d||d^{n}\alpha I).
\end{eqnarray}

\subsection{Two-level solution}

Our numerical calculations have shown that with a good accuracy the ground-state band and the first excited
negative parity band can be presented with a good accuracy as a superposition of two basis states of the form
(\ref{zerowf}). For the ground state band the two level approximation yields the wave
function of the form ($I$ can have only even values)

\begin{eqnarray}
\Psi^{g.s.}_I=&&\sin[\gamma_0(I)]\left [|\frac{I}{2}\frac{I}{2} I)\times Y_{0}
\right ]_{(IM)} - \cos[\gamma_0(I)] \left [ |\frac{I-2}{2}\frac{I-2}{2} (I-2))\times Y_{2}
\right ]_{(IM)},
\label{twolevel1}
\end{eqnarray}
where
\begin{eqnarray}
\sin[\gamma_0(I)]=&&\frac{1}{\sqrt{2}}
\left (
 1+\frac{1}{\sqrt{1+\left (\frac{V_0 \beta_0}{\sqrt{2\pi}\Delta}\right )^2 I}})
 \right )^{1/2}
 \end{eqnarray}
and for the energy we obtain
\begin{eqnarray}
\epsilon_I^{g.s.}=\hbar \omega \frac{I}{2}+\frac{\Delta}{2}\left [ 1 -
\sqrt{1+\left (\frac{V_0 \beta_0}{\sqrt{2\pi}\Delta}\right )^2 I} \right ].
\end{eqnarray}
In the last expressions $\Delta=\frac{3\hbar^2}{\mu R^2}-\hbar \omega$.

For the first excited negative parity bands we have
\begin{eqnarray}
\Psi^{n.p.}_I=&&\sin[\gamma_1(I)]\left [ |\frac{I-1}{2}\frac{I-1}{2} (I-1))\times Y_{1}
\right ]_{(IM)} + \cos[\gamma_1(I)] \left [ |\frac{I+1}{2}\frac{I+1}{2} (I+1))\times Y_{1}
\right ]_{(IM)},
\label{twolevel2}
\end{eqnarray}
where
\begin{eqnarray}
 \sin[\gamma_1(I)]=\frac{1}{\sqrt{2}}
\left (
 1+\frac{1}{\sqrt{1+
\left (\sqrt{\frac{12}{5\pi}}\frac{V_0 \beta_0}
{\hbar \omega} \right )^2\frac{(I+1)(2I+3)}{(2I+1)}}}
 \right )^{1/2}
\end{eqnarray}
with the energy is given as
\begin{eqnarray}
\epsilon_I^{n.p.}=\hbar \omega \frac{(I-1)}{2}+\frac{\hbar^2}{\Im}+
\hbar \omega
\left [1-\sqrt{1+
\left (\sqrt{\frac{12}{5\pi}}\frac{V_0 \beta_0}{\hbar \omega} \right )^2\frac{(I+1)(2I+3)}{(2I+1)}}
\right ].
\end{eqnarray}
Angular momentum $I$ can take only odd values.

\subsection{Multipole Moments}

Electric multipole operators are given by the expression

\begin{eqnarray}
\hat Q_{\lambda \mu} = \int \rho({\bf r})r^\lambda Y^\ast_{\lambda \mu}d\tau.
\end{eqnarray}

For the dinuclear system we assume that
\begin{eqnarray}
\rho({\bf r})=\rho_1({\bf r})+\rho_2({\bf r}),
\label{DNSdensity}
\end{eqnarray}
where $\rho_i \hspace{5pt}(i=1,2)$ are the densities of the DNS fragments.
Using (\ref{DNSdensity}) we can rewrite the expression for the electric multipole moments for
the DNS in the following form
\begin{eqnarray}
\hat Q_{\lambda \mu}=\sum_{\lambda_1,\lambda_1+\lambda_2=\lambda}
 \sqrt{\frac{4\pi(2\lambda+1)!}{(2\lambda_1+1)!(2\lambda_2+1)!}}
 \left [
 \hat q^{(\lambda_1 \lambda_2)}_{\lambda_1} \times Y_{\lambda_2}(\Omega)
 \right ]_{\lambda \mu},
 \label{multipole}
\end{eqnarray}
where
\begin{eqnarray}
\hat q^{(\lambda_1 \lambda_2)}_{\lambda_1}=
\left [
\left ( \frac{A_1}{A}\right )^{\lambda_2}Q_{\lambda_1}^{(2)}+
(-1)^{\lambda_2}\left ( \frac{A_2}{A}\right )^{\lambda_2}Q_{\lambda_1}^{(1)}
\right ]R^{\lambda_2}.
\end{eqnarray}
In the last expression
$Q^{(i)} \hspace{5pt} (i=1,2)$ are the intrinsic multipole moments of the DNS fragments.

Since we assume that the light fragment is spherical and can not be excited in the considered
energy range the only nonzero moment for the first fragment is $Q_{0}^{(1)}=Z_1/\sqrt {4\pi}$.
The second fragment is assumed to perform the quadrupole oscillations around the
spherical shape. Thus, in the linear approximation with respect to the deformation,
we have two nonzero moments for the second fragment $Q_{0}^{(2)}=Z_2/\sqrt {4\pi}$ and
$Q_{2}^{(2)}=\frac{3 Z_2 R_2^2}{4\pi}\beta^\ast_{2\mu}$.
Thus, we can write the expicit expressions for the dipole and quadrupole moment of the DNS
in the form
\begin{eqnarray}
Q_{1\mu}=
e\frac{A_1 Z_2-A_2 Z_1}{A}R \cdot Y_{1\mu}(\Omega)
\end{eqnarray}
for the dipole moment and
\begin{eqnarray}
Q_{2\mu}=
e\frac{A_1^2 Z_2+A_2^2 Z_1}{A^2}R^2 \cdot Y_{2\mu}(\Omega)+Q_{(2)}^{2 \mu}
\end{eqnarray}
for the quadrupole moment.

\subsection{Reduced transition probabilities}
Using  expression (\ref{wf}) for the wave function and (\ref{multipole}) for the multipole
operators we can calculate the reduced transition probabilities as the
\begin{eqnarray}
B(E\lambda,I_i \rightarrow I_f)=\frac{|<I_f||Q_{\lambda}||I_i>|^2}{2I_i+1}
\end{eqnarray}

The reduced matrix elements for the multipole operator $Q_{\lambda}$ between the initial state $i$ and the final state
$j$ has the following form
\begin{eqnarray}
&&<I_j p_j ||Q_{\lambda}||I_i p_i> \nonumber \\
&=&\sum_{\lambda_1 \lambda_2}\sum_{\{i\}\{j\}}a^{(I_j p_j)\ast}_{n_{I_1} \tau_{I_1}I_1 I_2}
a^{(I_i p_i)}_{n_{I'_1} \tau_{I'_1}I'_1 I'_2}
(n_{I_1} \tau_{I_1}I_1||q^{(\lambda_1 \lambda_2)}_{\lambda_1}||n_{I'_1} \tau_{I'_1}I'_1)C^{I_2 0}_{I'_2 0 \lambda_2 0}
\sqrt{\frac{(2\lambda+1)!}{(2\lambda_1+1)!(2\lambda_2+1)!}}\nonumber \\
&\times& \sqrt{(2i+1)(2j+1)(2\lambda+1)(2i_2+1)(2\lambda_2+1)}
\left \{\begin{tabular}{c c c}
$I_1$ & $I_2$ & $I_j$ \\
$I'_1$ & $I'_2$ & $I_i$ \\
$\lambda_1$ & $\lambda_2$ & $\lambda$
\end{tabular}
\right \},
\end{eqnarray}

where $\lambda_1=\lambda-\lambda_2$, and $\{i\}(\{j\})$ stands for the set of quantum numbers of the initial  (final)
states.

Using the two-level solutions for the ground-state band and the first excited negative parity bands we can easily calculate
the explicit expression for the intraband $B(E2)$ transitions and for the $B(E1)$  transitions between these bands.

In the case of the quadrupole transitions we have for the transitions between the states of the ground state band
 \begin{eqnarray}
 && B(E2,I^{g.s.}\rightarrow (I-2)^{g.s.}) = \nonumber \\
&& (-q_0^{(2,0)}\sin{[\gamma_0(I)]}\cos{[\gamma_0(I-2)]}+
q_2^{(0,2)}\sqrt{\frac{I}{2}}\sin{[\gamma_0(I)]}\sin{[\gamma_0(I-2)]}\nonumber \\
&& +
 q_2^{(0,2)}\sqrt{\frac{(I-2)}{2}}\cos{[\gamma_0(I)]}\cos{[\gamma_0(I-2)]})^2
 \end{eqnarray}
  and for the transition between the negative parity states
 \begin{eqnarray}
 && B(E2,I^{n.p.}\rightarrow (I-2)^{n.p.}) = \nonumber \\
&& (-q_0^{(2,0)}\sqrt{\frac{6(2I-3)}{5(2I-1)}}\sin{[\gamma_0(I)]}\cos{[\gamma_0(I-2)]}+
q_2^{(0,2)}\sqrt{\frac{I-1}{2}}\sin{[\gamma_0(I)]}\sin{[\gamma_0(I-2)]}\nonumber \\
&& +
 q_2^{(0,2)}\sqrt{\frac{(2I-3)(2I+3)(I+1)}{2(2I-1)(2I+1)}}\cos{[\gamma_0(I)]}\cos{[\gamma_0(I-2)]})^2.
 \end{eqnarray}
 In the last two expressions we have
\begin{eqnarray}
&& q_0^{(2,0)}=e_{eff}\frac{A_1^2 Z_2+A_2^2 Z_1}{A^2}R^2,\nonumber \\
&& q_2^{(0,2)}=e_{eff}\frac{3}{4\pi}Z_1 R_1^2\beta_0.
  \end{eqnarray}
For the dipole transitions between the two bands our calculations yields
\begin{eqnarray}
&& B(E1,I^{g.s.}\rightarrow (I-1)^{n.p.})\nonumber \\
&=&\frac{q_0^2}{2I+1}
\left \{
\sqrt{2 I-1}
\sin{[\gamma_0(I)]}\cos{[\gamma_1(I-1)]}+
\sqrt{\frac{6(2I+1)}{5}}\cos{[\gamma_0(L)]}\sin{[\gamma_1(I-1)]}
\right \}^2,
\nonumber \\
&& B(E1,I^{n.p.}\rightarrow (I-1)^{g.s.})
=q_0^2 \sin^2{[\gamma_1(I)]}\sin^2{[\gamma_0(I-1)]},
\label{BE1}
\end{eqnarray}
where
 $$q_0=e_{eff} \frac{A_1 Z_2-A_2 Z_1}{A} R$$.

\section{Results of calculations}

The results of  calculations of the energy spectra of
the ground state band and of the two lowest negative parity bands for the  $^{220}$Th are presented in Fig.2 together with the available experimental data. The calculation is done by the exact numerical diagonalization of the Hamiltonian (\ref{hamiltonian}). One can see the overall good agreement between the calculated and experimental spectra. The comparison of the experimental and calculated results for the states of the ground state  and of the first excited  negative parity bands are presented in Fig.3 in more details. It is seen  that the agreement is rather good at all values of angular momenta. Both, the ground-state and the first negative parity bands exhibit approximately an equidistant behavior which is related to the harmonic quadrupole oscillations of the heavy fragment.

\begin{figure}[t]
%\begin{right}
\includegraphics[bb=24 356 507 739, height=9.0 cm]{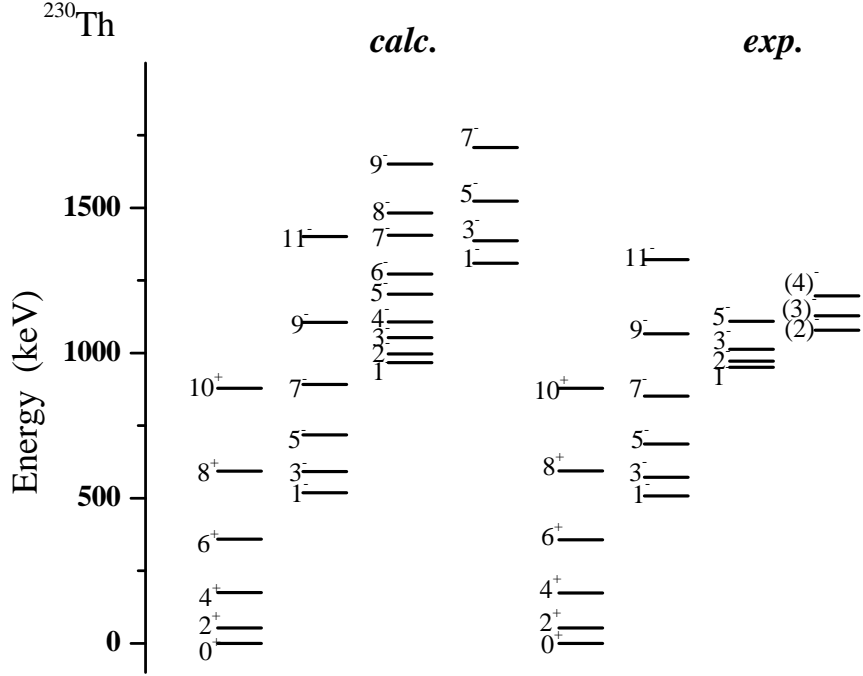}
%\end{right}
\begin{minipage}{16cm}
\vspace{0cm}
\caption{{\small \it
Calculated and experimental level scheme of $^{220}$Th. Experimental level, spin and parity assignments  are taken from \cite{reviol}.}}
\end{minipage}
\label{fig2}
\end{figure}

\begin{figure}[t]
%\begin{right}
\includegraphics[bb=77 408 542 705, height=9.0 cm]{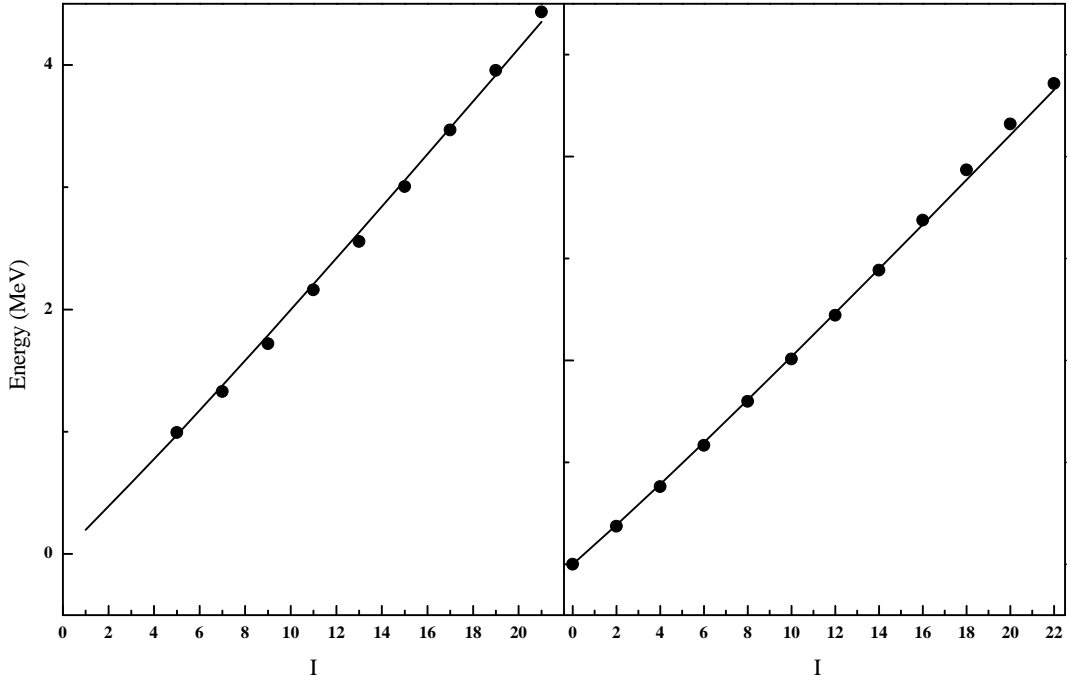}
%\end{right}
\begin{minipage}{16cm}
\vspace{0cm}
\caption{{\small \it
Calculated (lines) and experimental (solid circles connected by lines) energies of ground-state band (a) and first negative parity band (b). Experimental values are taken from \cite{reviol}.}}
\end{minipage}
\label{fig3}
\end{figure}

The calculation shows that mainly two eigenvectors of $\hat H_0$ are present in the wave function of the state of the ground state band, namely,   $\left [ |\frac{I}{2}\frac{I}{2} (I))\times Y_{0}\right ]_{(IM)}$  and
$\left [ |\frac{I-2}{2}\frac{I-2}{2} (I-2))\times Y_{2}\right ]_{(IM)}$. The contribution of the first of them is predominant at low angular momenta.
As a consequence at low angular momenta the ground state band  has an equidistant spectrum with the energy differences determined mainly by the frequency of the harmonic quadrupole oscillations of the heavy fragment. With increase of the angular momentum, the distance between the  levels is slightly increased, due to the growing admixture of the component
$\left [ |\frac{I-2}{2}\frac{I-2}{2} (I-2))\times Y_{2}\right ]_{(IM)}$ to the wave function. This introduces a small nonlinear dependence of the $\gamma$-transition energies as a function of the  angular momentum.

\begin{figure}[t]
%\begin{right}
\includegraphics[bb=62 407 542 705, height=8.0 cm]{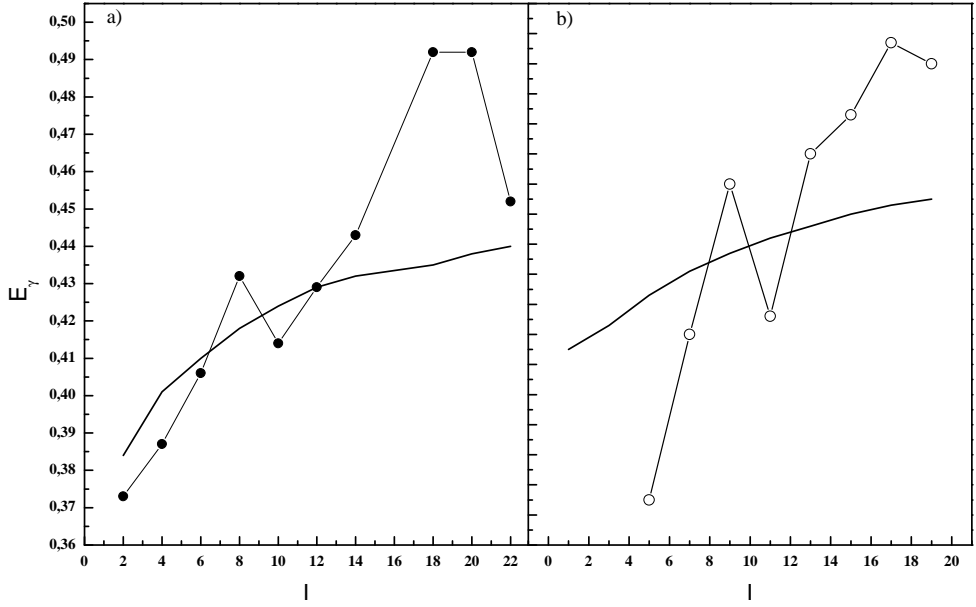}
%\end{right}
\begin{minipage}{16cm}
\vspace{0cm}
\caption{{\small \it
Calculated (line) and experimental (solid circles connected by lines) energies of $\gamma$ transitions between subsequent level of  the ground state band (a) and the first negative parity band (b). Experimental values  are taken from \cite{reviol}.}}
\end{minipage}
\label{fig4}
\end{figure}

The same equidistant structure with the frequency slightly growing with angular momentum holds for the first negative parity band. Again, the calculation shows that  mainly two eigenstates of $\hat H_0$  are present in the wave function. Namely,  $\left [ |\frac{I-1}{2}\frac{I-1}{2} (I-1))\times Y_{1}\right ]_{(IM)}$  and
$\left [ |\frac{I+1}{2}\frac{I+1}{2} (I+1))\times Y_{1}\right ]_{(IM)}$ . The contribution of the later component while being small at $I$=0 is  growing with angular momentum. The energy differences between the states in the negative parity band at low angular momentum are again determined mainly by the frequency of the quadrupole oscillations of the heavy fragment.

The angular momentum dependence of the vibrational frequency  $\omega_{vib}=E_\gamma/2$, defined as a half of the energy difference between the energies of two neighborhood levels of the  ground state  band and of the first excited negative parity band is  illustrated in Fig.4.  One can see the sharp decrease of the experimental values of $E_\gamma$ for the transition $10^+ \rightarrow 8^+$ in the ground state band and for the transition $13^-\rightarrow 11^-$ in the negative parity band which can be a consequence of the backbending phenomena. The backbending in these bands can be probably related to a rotational alignment of the nucleonic orbitals as it is mentioned in \cite{reviol}. The model considered above does not provide a  mechanism which could be responsible for the experimentally observed  behavior of the $\gamma$-transition energies. However, it is seen from Fig.~4 that the interval of change of  $\omega_{vib}$ with angular momentum observed  experimentally and obtained in calculations is not large. The energy differences between the neighborhood states in  bands are varying from  185 keV up to 240 keV. By taking into account the length of the spectra  in both energy and angular momentum, the frequency can be with good accuracy treated as a  constant.
\begin{figure}[t]
%\begin{right}
\includegraphics[bb=31 304 543 705, height=9.0 cm]{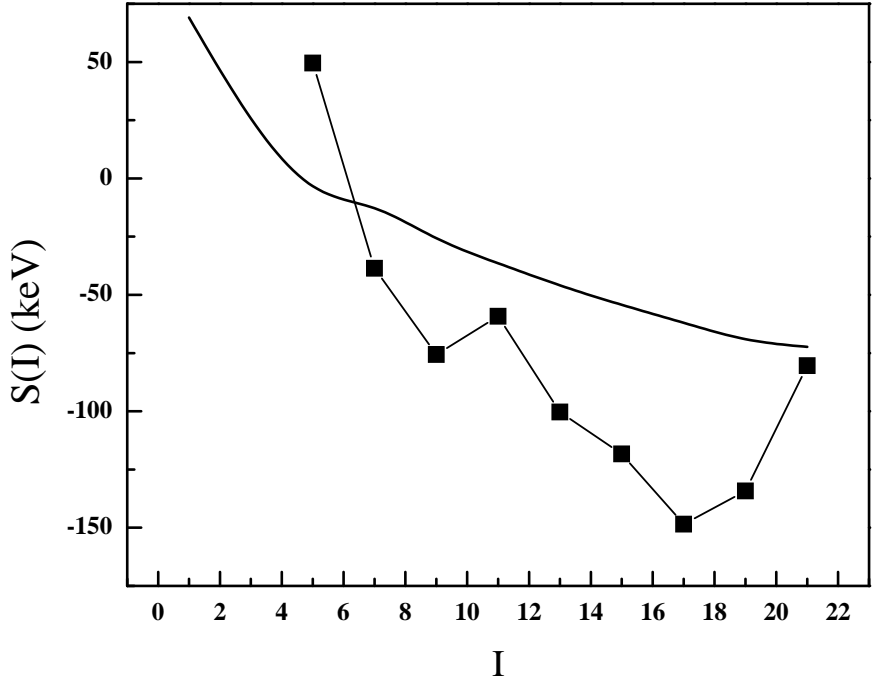}
%\end{right}
\begin{minipage}{16cm}
\vspace{0cm}
\caption{{\small \it
Calculated (lines) and experimental (solid squares connected by lines) values of parity splitting (see Eq.(\ref{psplitting})).
Experimental values are taken from \cite{reviol}.}}
\end{minipage}
\label{fig5}
\end{figure}

The dependence of experimental and calculated values of parity splitting in the ground and the first negative parity bands, treated as a unified alternating parity band, on angular momentum is illustrated in Fig.5.
The parity splitting is defined by the expression \cite{reviol}
\begin{eqnarray}
S(I^-)=E(I^-)-\frac{(I+1)E^+_{(I-1)}+I E^+_{(I+1)}}{2I+1}.
\label{psplitting}
\end{eqnarray}
It is seen from the figure that for the low angular momenta the parity splitting is positive and becomes negative with  angular momentum increase.  The possibility for the negative values of the parity splitting is related to the fact that the ground-state band and the first negative parity band are of the vibrational type. In this case the sign of the parity splitting is determined by the difference in  energies characterizing the quadrupole  vibrations of the heavy fragment and the rotation of the light fragment around the heavy one. Indeed, in the zero approximation $E(I^-)=\frac{1}{2}\omega (I-1)+\frac{\hbar^2}{2\mu R^2}$, $E(I^+)=\frac{1}{2}\omega I$ and therefore $S(I^-)=\frac{\hbar^2}{2\mu R^2}-\frac{1}{2}\omega \frac{2 I}{(2I+1)}$. Thus, for $\frac{\hbar^2}{2\mu R^2}<\frac{1}{2}\omega$, $S(I^-)$ can take negative values. In the case of a rotational bands, the value $S(I)$ must stay positive, achieving a zero value for the ideal unperturbed rotational bands of nucleus with stable octupole deformation.
\begin{figure}[t]
%\begin{right}
\includegraphics[bb=43 381 549 705, height=9.0 cm]{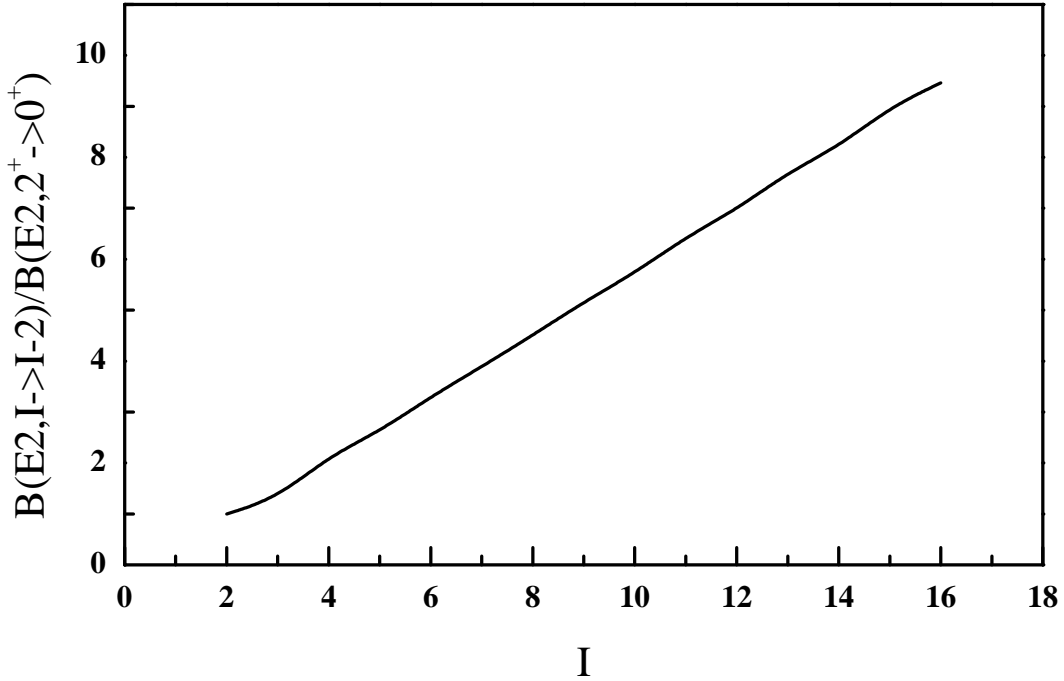}
%\end{right}
\begin{minipage}{16cm}
\vspace{0cm}
\caption{{\small \it
Angular momentum dependence of the ration of the reduced transition probabilities
$B(E2,I \rightarrow (I-2))/B(E2,2^+ \rightarrow 0^+)$ for the quadrupole transitions between the subsequent levels in the ground state band and the first negative parity band.}}
\end{minipage}
\label{fig6}
\end{figure}
The important feature of the spectra is an appearance at low energy of the second excited  negative  parity band which contains the states of even and odd angular momenta. The second excited negative parity band has an interesting features. The state with angular momentum $I$=2 is  lower than the state with $I$=1. With increase of angular momentum the normal level sequence is restored. The reason for such a behavior is related to the significant contribution of the states $\left [ |\frac{I+1}{2}\frac{I+1}{2} (I+1))\times Y_{1}\right ]_{(I+1, M)}$ and
$\left [ |\frac{I+1}{2}\frac{I+1}{2} (I+1))\times Y_{1}\right ]_{(I+2, M)}$ in the wave functions for even and odd angular momenta, respectively. In the limit if $\hat V_{int}$ is equal to zero, these two states are degenerate.

In Fig.6, the values of the ratio $B(E2,I\rightarrow I-2)/B(E2,2^+ \rightarrow 0^+)$ are presented as a function of the initial angular momentum. As it should be in the case of harmonic quadrupole oscillations of the heavy fragment  the values of $B(E2)$  increase linearly with $I$. They does not show any changes in the behavior for  transitions between the members of the ground-state  band and the first negative parity band since the underlying quadrupole constituents in both bands are the same. Thus, the experimentally observed staggering of $BE1/BE2$ ratios can be attributed to the staggering of the $B(E1)$ values (see Fig.~7).

\begin{figure}[t]
%\begin{right}
\includegraphics[bb=49 306 543 705, height=9.0 cm]{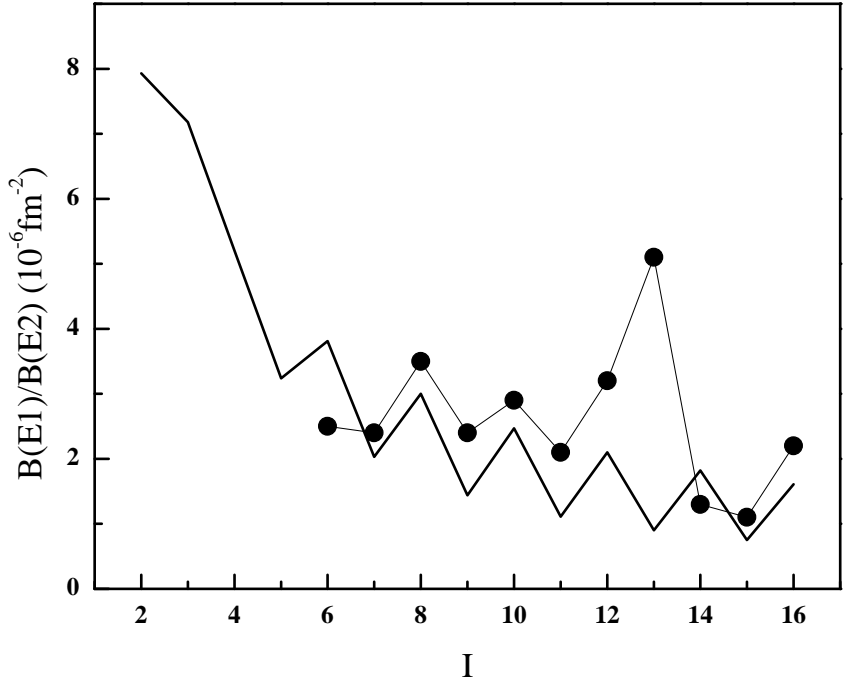}
%\end{right}
\begin{minipage}{16cm}
\vspace{0cm}
\caption{{\small \it
 $B(E1)/B(E2)$ ratios as a function of the initial angular momentum for transitions in the yrast and first negative parity bands. Experimental values (filled circles) are taken from \cite{reviol}.}}
\end{minipage}
\label{fig7}
\end{figure}

 Such a staggering behavior of $B(E1)$ can be qualitatively explained with the use of
equation (\ref{BE1}). We can see that the reduced transition probability B(E1) for the transition from state $I$ of the ground state  band to the state ($I$-1) of the first excited negative parity band consist of two contributions, since  both dipole transitions are allowed (see Eqs.(\ref{twolevel1},\ref{twolevel2})) from the component  $\left [ |\frac{I}{2}\frac{I}{2} (I))\times Y_{0}\right ]_{(IM)}$ to
$\left [ |\frac{I}{2}\frac{I}{2} (I))\times Y_{1}\right ]_{(I-1,M)}$ and from $\left [ |\frac{I-2}{2}\frac{I-2}{2} (I-2))\times Y_{2}\right ]_{(IM)}$ to $\left [ |\frac{I-2}{2}\frac{I-2}{2} (I-2))\times Y_{1}\right ]_{(I-1,M)}$.
In the opposite case of transition from the states of the negative parity band to the states of the positive parity belonging to the ground state band, we have only one allowed transition, namely from the component $\left [ |\frac{I}{2}\frac{I}{2} (I))\times Y_{1}\right ]_{(I+1,M)}$ to the component
$\left [ |\frac{I}{2}\frac{I}{2} (I))\times Y_{0}\right ]_{(I,M)}$. The transition from
$\left [ |\frac{I+2}{2}\frac{I+2}{2} (I+2))\times Y_{1}\right ]_{(I+1,M)}$ to
$\left [ |\frac{I-2}{2}\frac{I-2}{2} (I-2))\times Y_{2}\right ]_{(I,M)}$ is forbidden because the dipole operator  does not change a number of the quadrupole phonons.

 The $B(E1)/B(E2)$ ratios as a functions of an initial angular momentum are presented in Fig.7.  Calculated ratios for the odd initial angular momentum (i.e. for transitions from the states of the negative parity) lie systematically lower than the ratios for the even initial angular momentum (transitions from the state of the ground state band). This is in agreement with the experimental data with the exception of two data points at $13^-$ and $14^+$ .  As it is mentioned in \cite{reviol} the large value of $B(E1)/B(E2)$ ratio at  $13^-$ can be attributed to the loss of $E2$ strength in the backbending. The rather small  $B(E1)/B(E2)$ value for the  $14^+$
 attributed to the spread of $E1$ strength due to the presence of two $13^-$ final states.

\section{CONCLUSION}

We suggest a cluster interpretation of the multiple negative parity bands in $^{220}$Th assuming  collective oscillations of nucleus in mass-asymmetry degree of freedom.  This collective motion leads to the admixture of the very asymmetric cluster configurations to the intrinsic nucleus wave function and creates deformations with even- and odd-multipolarities. To take care of reflection asymmetric modes with nonzero values of $K$, the harmonic quadrupole oscillations of the heavy cluster is considered. The resulting energy spectrum consists of the ground state band and several negative parity bands which exhibit nearly equidistant behavior. The angular momentum dependence of the parity splitting is described. The possibility for the  negative values of the parity splitting is related in the model to the interplay between the quadrupole vibrations of the heavy fragment and the rotational motion of the light fragment.
We described the staggering behavior of the $B(E1)/B(E2)$-rations as a function of the angular momentum. The $BE(1)$ transitions from the state of the negative parity to the state of the positive parity is hindered, because these states contain  different numbers of the quadrupole phonons. The results of calculations are in overall good agreement with experimental data. This work is further development of the previously developed approaches \cite{shneyd,shneyd1,shneyd2}.

\end{document}